\documentclass[aps,pre,reprint,superscriptaddress,longbibliography]{revtex4-1}
\usepackage{graphicx}
\usepackage{hyperref}
\usepackage[utf8]{inputenc}

\usepackage{color}

\begin{document}

\title{Coupling of link- and node-ordering in the coevolving  voter model}

\author{J. Toruniewska}
\email[]{toruniewska@if.pw.edu.pl}
\affiliation{ Center of Excellence for Complex Systems Research,\\ Faculty of Physics, Warsaw University of Technology,\\ ul. Koszykowa 75, PL-00662 Warsaw, Poland}
\author{K. Ku{\l}akowski}
\email[]{kulakowski@fis.agh.edu.pl}
\affiliation{Faculty of Physics and Applied Computer Science, AGH University of Science and Technology,\\ al. Mickiewicza 30, PL-30059 Krak\'ow, Poland}
\author{K. Suchecki}
\email[]{suchecki@if.pw.edu.pl}
\affiliation{ Center of Excellence for Complex Systems Research,\\ Faculty of Physics, Warsaw University of Technology,\\ ul. Koszykowa 75, PL-00662 Warsaw, Poland}
\author{J. A. Ho{\l}yst}
\email[]{jholyst@if.pw.edu.pl}
\affiliation{ Center of Excellence for Complex Systems Research,\\ Faculty of Physics, Warsaw University of Technology,\\ ul. Koszykowa 75, PL-00662 Warsaw, Poland}
\affiliation{ITMO University, 19 Kronverkskiy av., 197101 Saint Petersburg, Russia}
\affiliation{Netherlands Institute for Advanced Study in the Humanities and Social Sciences, PO Box 10855, 1001 EW Amsterdam, The Netherlands}

\date{\today}

\begin{abstract} 
We consider the process of reaching the final state in the coevolving voter model.  There is a coevolution of state dynamics, where a node can copy a state from a random neighbor with probabilty $1-p$ and link dynamics, where a node can re-wire its  link to another node of the same state with probability $p$. That exhibits an absorbing transition to a frozen phase above a critical value of rewiring probability. Our analytical and numerical studies show that in the active phase mean values of magnetization of nodes $n$ and links $m$ tend to the same value that depends on initial conditions. In a similar way mean degrees of spins up and spins down become equal. The system obeys a special statistical conservation law since a linear combination of both types magnetizations averaged over many realizations starting from the same initial conditions is a constant of motion: $\Lambda\equiv (1-p)\mu m(t)+pn(t) = const$, where $\mu$ is the mean node degree.  The final mean magnetization of nodes and links in the active phase is proportional to  $\Lambda$   while the final density of active links is a square function of  $\Lambda$.  If the rewiring probability is above a critical value and the system separates into disconnected domains, then the values of nodes and links magnetizations are not the same and final mean degrees of spins up and spins down can be different. 
 
\end{abstract}

\maketitle

\section{Introduction}

 It is common that a system in an equilibrium state is described by appropriate  balance equations, that possess mechanical, thermal, chemical, or other meaning   \cite{Kondepudi}. It is no different when we move toward a multi-agent model with stochastic dynamics \cite{Gardiner},\cite{Weidlich}. The aim of this paper is to find  corresponding  balance conditions for a coevolving voter model \cite{ref_voter_coevolution}.

The model was introduced as a simple model of competition between species \cite{ref_invasion} but later named voter model \cite{ref_voter_math}.
Its simplicity means it could be used in many contexts \cite{ref_castellano_review}, including opinion formation or catalytic reactions \cite{ref_voter_krapivsky}.
The basic two-state voter model has several noteworthy properties: the coarsening of domains leading to ordering that depends on dimensionality \cite{ref_voter_krapivsky}, lack of surface tension on domain boundaries \cite{ref_voter_tension}, as well as statistical conservation of the number of different states \cite{ref_voter_conservation}.
The last property is conditional on whether one updates nodes or links, or viewing differently, how the source and target nodes for the state overwriting are chosen \cite{ref_voter_conservation}.
It is also statistical, which means that the conservation happens only for averages over many realizations, while each single realization will inevitably change the numbers of nodes in different states, leading at the end to an absorbing, fully ordered state of a finite system.
The statistical nature makes these conservation laws different than conservation of energy or momentum, that are strictly fulfilled.
The voter model was exhaustively researched \cite{ref_castellano_review} both in finite-dimensional systems \cite{ref_voter_cluster,ref_voter_krapivsky,ref_voter_sierpinski} as well on networks, including random graphs and scale free networks \cite{ref_voter_glaubercompare,ref_voter_sood,ref_voter_uncorrelated}, Watts-Strogatz small-world networks \cite{ref_voter_smallworld}, networks with modular structure \cite{ref_voter_modular}, and directed networks \cite{ref_voter_directed,ref_voter_conservationdirected}.
Variations of the original voter model were also studied, including noisy voter model \cite{ref_voter_noisy,ref_voter_noisycomplex}, nonlinear voter model \cite{ref_voter_nonlinear,ref_qvoter} as well as a few other \cite{ref_castellano_review}.

One of more interesting variants of the voter model is the case of coevolution of the voter dynamics and network topology.
The coevolution dynamics is also known as adaptive networks \cite{ref_adaptive}.
In the nontrivial situation, where the topology of the network changes in response to the voter dynamics on the network with similar times cales, new phenomena arise.
The best known is the fragmentation transition \cite{ref_fragmentation, ref_voter_coevolution} where the coevolution can lead to a partition of the network into several separate clusters, in the case of voter model each with internal ordering of node states.
This transition has been also studied in more complex network types, including directed \cite{ref_voter_fragmentationdirected} and multi-layer networks \cite{ref_voter_fragmentationmultiplex}.
The issue of dynamics of so-called link magnetization was considered in \cite{ref_fragmentation, ref_voter_uncorrelated}, allowing to calculate how the system approaches a final absorbing state of a static network and when the fragmentation transition occurs in coevolutionary dynamics. While the coevolving voter model was studied, the research mostly focused on why and how the fragmentation occurs, and while works have studied the dynamics of numbers of links connecting different state combinations \cite{ref_voter_fragmentationanalytical,ref_voter_coevolution,ref_voter_reverse}, they have not focused on conservation laws.

In this paper we investigate the coevolution of the voter model \cite{ref_invasion,ref_voter_math,ref_voter_krapivsky} and network topology, where links can be re-wired \cite{ref_voter_fragmentationanalytical} to connect to another node of the same state instead of changing the state of a node. We explore  relations between mean magnetization of links and nodes. In our mean-field calculations, we treat mean degrees of nodes in different states $(+)(-)$ as separate variables $\mu_+,\mu_-$. This allows discussing the magnetization of links and the magnetization of nodes as potentially independent variables. Yet we find that a combination of node and link magnetizations is conserved in an ensemble average (Eq. \ref{cofmo}), which is consistent with the conservation of weighted spins for non-adaptive networks \cite{ref_voter_conservation} as well as an obvious fact that node magnetization is conserved when only links can change.  Moreover, we show that when the rewiring probability is below the critical value and the final network state contains a nonzero value of active links then the mean magnetizations of nodes and links are the same and mean degrees of nodes possessing different spins also become equal.

\section{Elementary events}

Let us consider a complex network where every node is attributed to an internal variable that will be  called a spin and is valued $+1$   or $-1$ (up or down).  The variable can correspond also to two different opinions of agents placed in network nodes. Let $N_+$, $N_-$ be numbers of nodes up and down, respectively, and the total number of nodes is $N_+ + N_-=N$. We do not specify the network topology that will evolve in  the course of time. Suppose that each link between nodes is cut in half and the half-links are classified as directed. Their number is denoted here as $M_{\alpha \beta}$ (from the nodes endowed with spin $\alpha$ to the nodes endowed with spin $\beta$), where $\alpha, \beta =\pm 1$. Obviously, $M_{+-}=M_{-+}$, and $M_{++}+M_{--}+2M_{+-}=N\mu$, where $\mu$ is the mean degree of node in the network. Further, a node with the spin $\alpha$ is supposed to have $k_\alpha=k_{\alpha\alpha}+k_{\alpha\beta}$  neighbors ($\beta \neq \alpha$), where $k_{\alpha\alpha}$ ($k_{\alpha\beta}$) is the number of half-links directed from the node with spin $\alpha$ to other nodes with spins $\alpha$ ($\beta$), and so on.\\

The system dynamics consists of the evolution of spins as in the standard voter model and changes in  network topology as was suggested in   \cite{ref_voter_fragmentationanalytical}. In every time step a random node $i$  is chosen and then  one of its nearest neighbors $j$ is selected. If both nodes possess the same spins the connecting link will be called {\it inert} and nothing happens. If spins are different then the connecting link will be called {\it active} and there are two options.   \begin{itemize}
\item With a probability $r=1-p$ the spin $i$  is changed to the value of the spin $j$. It means the  agent $i$ has been  convinced by one of his neighbors $j$  to change his opinion. Such a case corresponds to the standard voter dynamics \cite{ref_invasion,ref_voter_math,ref_voter_krapivsky}. 
\item With a probability $p$ the spin  $i$ is kept but a link between nodes $i$ and $j$ is removed and a new link between the node $i$ and any other node $l$ that was initially disconnected from $i$ but possessed the same spin as the node $i$ is created \cite{ref_voter_fragmentationanalytical}. Using  the language of opinion dynamics such a rewiring process can correspond to the case when the agent $i$ was not able to accept the opinion of the agent $j$ and this disagreement leads to breaking by the agent $i$ his social relationship to the agent $j$  and replacing it with a new social tie to the agent $l$ that shares the same opinion as the agent $i$. 
\end{itemize}

Overall, there are four possible elementary events: state changes $+ \to -$ and $- \to +$ as well as rewirings $+- \to ++$ and $-+ \to --$. Each of these events changes numbers of spins of given state $N_+$ and $N_-$ as well as numbers of links connecting given states $M_{++}$,$M_{+-}$,$M_{-+}$,$M_{--}$ in the following ways (symbols $k_{\alpha\beta}$ refer to links from the updated node $\alpha$ to a node $\beta$):\\

$\it (flips)$ $+ \to -$: $N_+ \to N_+-1$, $N_- \to N_-+1$,  $M_{++}\to M_{++}-2k_{++}$,  $M_{--}\to M_{--}+2k_{+-}$, $M_{+-}\to M_{+-}+k_{++}-k_{+-}$ and $M_{-+} \to M_{-+}+k_{++}-k_{+-}$,\\

$\it (flips)$ $- \to +$: exactly opposite to above,\\

$\it (rewirings)$ $+- \to ++$: $M_{++}\to M_{++}+2$, $M_{+-}\to M_{+-}-1$ and $M_{-+}\to M_{-+}-1$,\\

$\it (rewirings)$ $-+ \to --$: exactly opposite to above.\\

\begin{figure}[!hptb]
\begin{center}
\includegraphics[width=.8\columnwidth, angle=0]{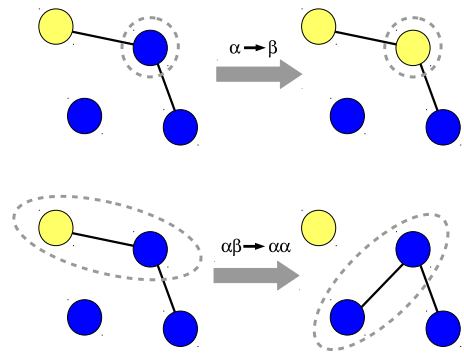}
\caption{(Color online) Possible elementary events:  flip $\alpha \to \beta$ (upper) and rewiring $\alpha\beta \to \alpha\alpha$ (bottom).}
\label{events}
\end{center}
\end{figure}

These rules are illustrated by Fig. \ref{events}. For example, a flip $+ \to -$ means that $k_{++}$ inert links of a given node become active links of type $-+$ and as result the variable $M_{-+}$ increases by $k_{++}$. Simultaneously, $k_{+-}$ active links of this node are switched to inert links of type $--$, and so on.

Let us note that flips of spins and rewiring of links keep the total number of links in the network constant.

\section{Mean-field calculations}
Now mean-field  equations of motion will be  constructed in a similar way as it was done in \cite{ref_voter_coevolution}.   We shall assume, however, that during the system evolution mean degrees of nodes $\mu_{\alpha} = \langle k_{\alpha}\rangle$ possessing positive and negative spins can be different, similarly fractions of active links around positive and negative spins can be also different. As far as we know, this approach has never been applied before.  We shall show that these more general assumptions lead to time dependence of mean magnetization of nodes and mean magnetization of links as well as to nontrivial changes in dynamics of active links. Let $P_\alpha(k_\alpha)$ be the degree distribution for nodes endowed with spin $\alpha$ and $B(k_{\alpha \beta};k_\alpha)$ be the probability of $k_{\alpha\beta}$ active links out of $k_\alpha$ links around nodes endowed with spin $\alpha$. A general equation of motion for the total number of active links can be written as 

\begin{eqnarray}
 &\frac{dM_{\alpha\beta}}{dt}=N_\alpha\sum_{k_\alpha} P_\alpha(k_\alpha) \sum_{k_{\alpha\beta}} B(k_{\alpha\beta};k_\alpha)\nonumber \\ &\cdot \frac{k_{\alpha\beta}}{k_\alpha}[r(k_{\alpha\alpha}-k_{\alpha\beta})-p]+N_\beta\sum_{k_\beta} P_\beta(k_\beta) \nonumber \\ &\cdot \sum_{k_{\beta\alpha}} B(k_{\beta\alpha};k_\beta) \frac{k_{\beta\alpha}}{k_\beta}[r(k_{\beta\beta}-k_{\beta\alpha})-p].
 \label {eq1}
\end{eqnarray}

Dynamics of the spin numbers $N_\alpha$ is 

\begin{eqnarray}
\frac{d N_{\alpha}}{dt}=\frac{-r}{1/N_{\alpha}} \sum_{k_{\alpha}} P_{\alpha}(k_{\alpha}) \sum_{k_{\alpha \beta}} B(k_{\alpha \beta};k_{\alpha})\frac{k_{\alpha \beta}}{k_{\alpha}}+\nonumber \\
+\frac{+r}{1/N_{\beta}} \sum_{k_{\beta}} P_{\beta}(k_{\beta}) \sum_{k_{\beta \alpha}} B(k_{\beta \alpha};k_{\beta})\frac{k_{\beta \alpha}}{k_{\beta}}.
 \label {eqNa}
\end{eqnarray}

The above equations follow directly from the description of elementary events in our dynamics.
We assume that the probability distribution $B(k_{\alpha\beta};k_\alpha)$  is binomial, treating it as if a result of $k_\alpha$ trials and with a success rate $\eta_\alpha$ equal the fraction of active links around nodes endowed with spin $\alpha$.
Let us stress that $k_{\alpha}$ can be a time-dependent variable, and in general it is possible that $\eta_+ \neq \eta_-$ and  $\mu_+ \neq \mu_-$  where $\mu_\alpha$ is the mean degree $\langle k_\alpha \rangle$. In this point our model is more general than that considered, for example, in \cite{ref_voter_coevolution} where only symmetrical states $\mu_+=\mu_-=\mu$ and $\eta_+=\eta_-$ were taken into account.

The first two moments of the distribution  $B(k_{\alpha\beta};k_\alpha)$ can be easily found   
\begin{eqnarray}
\sum_{k_{\alpha \beta}} B(k_{\alpha \beta};k_{\alpha})k_{\alpha \beta   }&=&k_{\alpha}\eta_\alpha, \nonumber\\
\sum_{k_{\alpha  \beta}} B(k_{\alpha   \beta };k_{\alpha})k_{\alpha   \beta}^2&=&k_{\alpha}^2\eta_\alpha^2+k_{\alpha}\eta_\alpha(1-\eta_\alpha).\nonumber\\
 \label{mf}
\end{eqnarray}

Using these relations and taking into account that $k_\alpha=k_{\alpha \alpha}+ k_{\alpha \beta}$ after some algebra we get from Eq. (\ref{eq1}) \\

\begin{eqnarray}
 \frac{dM_{\alpha \beta}}{dt}&= N_\alpha[r\eta_\alpha \mu_\alpha-2r\eta_\alpha^2\mu_\alpha-2r\eta_\alpha(1-\eta_\alpha)-p\eta_\alpha]+\nonumber \\
 &+N_\beta[r\eta_\beta \mu_\beta-2r\eta_\beta^2\mu_\beta-2r\eta_\beta(1-\eta_\beta)-p\eta_\beta],
\label{eq2}
\end{eqnarray}

 Similarly, 

\begin{eqnarray}
 \frac{dM_{\alpha  \alpha}}{dt}= -2N_\alpha r \eta_\alpha(\mu_\alpha-\mu_\alpha\eta_\alpha-1+\eta_\alpha)+\nonumber\\
 +2N_\alpha p\eta_\alpha+2rN_\beta\eta_\beta(\mu_\beta\eta_\beta+1-\eta_\beta), 
 \label{eq3}
\end{eqnarray}

\begin{eqnarray}
\frac{dN_{\alpha}}{dt}=r(N_{\beta}\eta_{\beta}- N_{\alpha}\eta_{\alpha}).
\end{eqnarray}

To make the notation more convenient, we define the density $\rho$ of active links by its relation with $M_{+-}$, i.e. $M_{+-}=M_{-+}=N\mu \rho /2$. Also, we introduce order parameters for mean magnetizations of nodes $n$ and links $m$, defined by the relations: $N_+-N_-=Nn$, and $M_{++}-M_{--}=N\mu m$. The link magnetization $m$ is equal to the weighted magnetization described in \cite{ref_voter_conservation}. Hence, $N_\alpha=N(1+\alpha n)/2$, and recalling that $M_{\alpha \alpha}+ M_{\beta \beta}+2M_{\alpha \beta }=N\mu$ we have $M_{\alpha \alpha}=N\mu(1-\rho+\alpha m)/2$. Further, from the definition of the mean degrees of nodes   $\mu_{\alpha}$ we have $N_{\alpha}\mu_{\alpha}=M_{\alpha \alpha}+ M_{\alpha \beta}$. Then,
\begin{equation}
\mu_\alpha = \frac{\mu(1+\alpha m)}{(1+\alpha n)}.
\label{eq_mua}
\end{equation}
Finally, the coefficients $\eta_{\alpha}$ can be written as $\eta_{\alpha}=M_{\alpha \beta}/(N_{\alpha}\mu_{\alpha})$, hence  $\eta_{\alpha}=\rho/(1+\alpha m)$.

One can see that $\eta_+=\eta_-$ only when $m=0$ or $\rho=0$. Similarly the mean degrees of nodes with spin $\alpha$ are the same for $m=n$. We shall consider a triple $(\rho, m, n)$ as a set of  time-dependent observables describing our system. Equations of motion for these new variables are

\begin{eqnarray}
 \frac{d\rho}{dt}=&\frac{2r\rho}{1-m^2}[1-m^2-2\rho-\frac{2}{\mu}(1-m n)+\frac{2\rho(1+m^2-2nm)}{\mu(1-m^2)}]\nonumber \\
 &-\frac{2p\rho(1-m n)}{\mu(1-m^2)},
 \label{derodete}
\end{eqnarray}

\begin{equation}
 \frac{dn}{dt}=\frac{2r\rho(m-n)}{1-m^2},
 \label{deemdete}
\end{equation}
and
\begin{equation}
 \frac{dm}{dt}=\frac{2p\rho(n-m)}{\mu(1-m^2)}.
 \label{debetadete}
\end{equation}
If $m=0$, Eq. (\ref{derodete}) reduces to Eq. (2) in \cite{ref_voter_coevolution}. On the other hand, Eqs. (\ref{deemdete}) and (\ref{debetadete}) indicate that the order parameters $n$ and $m$ are coupled. Eliminating the explicit time dependence, we receive a new conservation law for a linear combination of both magnetizations 

\begin{equation}
 (1-p)\mu m(t)+pn(t)=\Lambda,
 \label{cofmo}
\end{equation}
where $\Lambda$ is a  constant of motion for this system that results from initial conditions $m_0=m(t=0)$ and $n_0=n(t=0)$. Let us remark  that the conservation law (\ref{cofmo}) is fulfilled only statistically. Every elementary update (flip or rewiring) changes $m$ and $n$ in a way that $\Lambda_R$ (the value for the actual realization calculated from Eq.(\ref{cofmo}) with real $m(t)$ and $n(t)$) is also changing. However, due to the symmetry of probabilities $\Lambda_R$ is only experiencing fluctuations similar to unbiased random walk.
Since  $|m| \le 1$ and $|n| \le 1$ thus $|\Lambda| \le  (1-p)\mu +p$.
In the limiting case for $p=1$ (no flips), we get $n=const$, what is obvious. On the other hand, for $p=0$ (no rewiring) we get $m=const$. This result can be understood as follows. 
A state change can only follow from  interaction along any of the active links (between nodes $i$ and $j$ possessing different spins $s_i=+1$ and $s_j=-1$). This is either the  flip $+ \to -$ (probability $1/(Nk_+^i))$ or the flip  $- \to +$ (probability $1/(Nk_-^j)$. The related changes of $M_{++}-M_{--}$ are $-2(k_{++}^i+k_{+-}^i)$ or $2(k_{-+}^j+k_{--}^j)$, respectively. As $k_{++}^i+k_{+-}^i=k_+^i$ and $k_{-+}^j+k_{--}^j=k_-^j$, the mean change of $M_{++}-M_{--}$ is zero. A similar argumentation on the preservation of magnetization weighted by the node degree was presented in \cite{ref_voter_conservation}. In this way our relation (\ref{cofmo}) links together two distinct conservation laws that are fulfilled in  the limiting cases $p=0$ and $p=1$.     \\
Another conclusion from Eqs. (\ref{deemdete}) and (\ref{debetadete}) is the that asymptotic values of the links and nodes magnetizations $m$ and $n$ are equal 
\begin{equation} 
m^*=n^*
\end{equation} 
provided that the asymptotic density of active links is nonzero $\rho^*>0$. In other words, active links are responsible for reaching the balance between the links and nodes magnetization.  
As a consequence of Eq. (\ref{cofmo}), the number of variables is reduced to two; let us take ($\rho$, $m$). The evolution of the variable $n(t)$ is given by Eq. (\ref{cofmo}), or explicitly through initial conditions as
\begin{equation}
 n(t)=n_0+\mu \frac{1-p}{p}(m_0-m(t)).
\label{modbeta}
\end{equation}
Equations \ref{derodete} and \ref{debetadete} with $n$ calculated from Eq. (\ref{modbeta}) have a line of fixed points $(\rho=0,m)$ and a fixed point $(\rho^*,m^*)$ which depends on $\Lambda$ and thus the initial state
\begin{eqnarray}
\rho^* &=& \left[ 1 - \left(\frac{\Lambda}{\mu (1-p) + p}\right)^2 \right] \frac{(1-p)(\mu-1)-1}{2(1-p)(\mu-1)} \label{fxpt1} \\
m^* &=& \frac{\Lambda}{\mu (1-p) + p}. \label{fxpt}
\end{eqnarray}
The last equation shows that the constant of motion  $\Lambda$ can be seen as the final value of the node or link magnetization normalized by a linear combination of system parameters $\mu$ and $p$. Following Eq.(\ref{fxpt1}) the final density   $\rho^*$  is a  quadratic function of $\Lambda$, it reaches its  maximal value for  $\Lambda=0$ and it  vanishes when  $|\Lambda|$ is maximal.    
\begin{figure}[!hptb]
\begin{center}
\includegraphics[width=\columnwidth, angle=0]{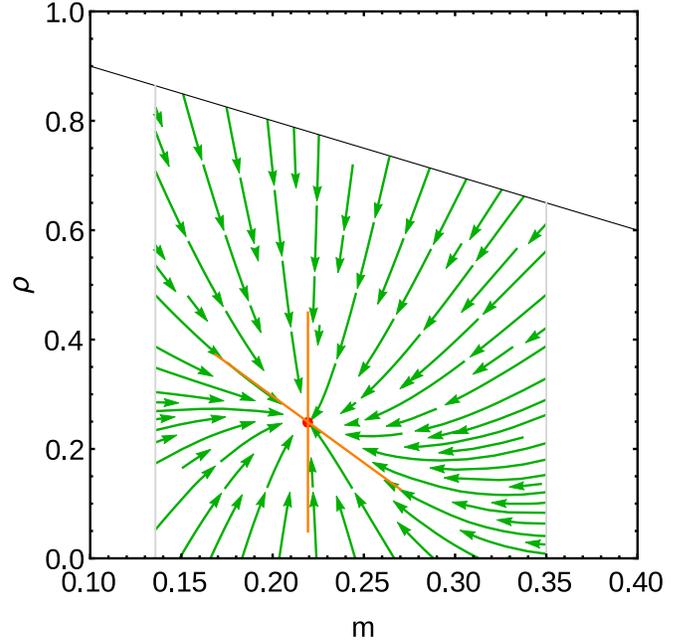}
\caption{(Color online) Expected flows of the system state $(m,\rho)$ in phase space according to Eq.\ref{derodete} to \ref{debetadete} with $n$ eliminated by Eq. \ref{modbeta}. Green lines represent direction of state changes (speed is not indicated in any way), red dot shows stable fixed point $(\rho^*,m^*)$ expressed by Eqs.\ref{fxpt1}, \ref{fxpt} and orange crossing lines show the directions of the eigenvectors of the Jacobian at the fixed point. Figure uses $p=0.3$, $\mu=4$, $\Lambda=0.68$ (this $\Lambda$ includes for example $m_0=0.2$, $n_0=0.4$). Note that since $n_0 \in (-1,1)$, thus Eq.\ref{cofmo} places limits on possible $m$ for given $\Lambda$. Grey lines show limits of possible $m$ and black line shows limit of $\rho$ possible for given $m$.}
\label{voterkk_flowstab}
\end{center}
\end{figure}
The density $\rho^*$ can be also expressed as a function of $m^*$ instead of $\Lambda$, where the whole dependence on initial conditions is via the value of $m^*(\mu,p,n_0,m_0)$
\begin{equation} \label{rhostar1}
\rho^*=(1-m^{*2})\frac{(1-p)(\mu-1)-1}{2(1-p)(\mu-1)}.
\end{equation}
Since in the stationary state $m^*=n^*$, this relation is the same as the one observed in numerical simulation results in \cite{ref_voter_coevolution}, where the trajectories of model realizations were observed to fluctuate along such a curve.\\

The fixed point Eq.(\ref{fxpt1}) and (\ref{fxpt}) depends on both parameters $p,\mu$, and $\Lambda(p,\mu,n_0,m_0)$. The stability analysis shows that parameter $p$ alone decides whether it is stable or not. The Jacobian for the fixed point $(\rho^*,\mu^*)$ has eigenvalues
\begin{eqnarray}
\lambda_1 &=& \frac{\left[ \mu(1-p)+p \right] \left[ 2-\mu(1-p)-p \right]}{(\mu-1)\mu(1-p)} \label{eqlambda1} \\
\lambda_2 &=& \frac{2}{\mu}\left[ 2-\mu(1-p)-p \right] \label{eqlambda2}
\end{eqnarray}
with associated eigenvectors: $u_1=[ \rho_1, m_1 ]$ and $u_2=[1,0]$,
where:
\begin{eqnarray}
\frac{\rho_1}{m_1} = m^* \cdot \frac{2\mu + 3(\mu-2)(\mu-1)p - 3(\mu-1)^2 p^2}{p (\mu-1) (2-\mu (1-p)-p)}
\label{rho_over_m}
\end{eqnarray}
For $p \to p_c$, the denominator of Eq. (\ref{rho_over_m}) becomes zero, so the second component of the eigenvector $u_1$ becomes zero and the eigenvectors are parallel.
Since  
\begin{equation}\label{l1l2}
\lambda_1=\lambda_2\frac{\mu(1-p)+p}{2(\mu-1)(1-p)} 
\end{equation}
thus both eigenvalues in the limit $p \to p_c$, where 
\begin{equation}\label{p_c}
p_c(\mu)=\frac{\mu-2}{\mu-1}
\end{equation}
are zero [follows Eqs. (\ref{eqlambda1}) and (\ref{eqlambda2})] and are negative as functions of the probability $p$ below the same point.
It follows the fixed point is stable for $p<p_c$ and unstable for $p>p_c$.
The value of $p_c$ is the same as the critical threshold of the transition to the frozen phase found in \cite{ref_voter_coevolution}.\\
 For $p>p_c(\mu)$ the point $(\rho^*, m^*)$ not only becomes unstable, but it is also outside the phase space of the system ($\rho^*<0$) and thus cannot be ever reached.
Figure \ref{voterkk_flowstab} shows an example of the flows of the system state $(\rho,m)$ in phase space, along with fixed point $(\rho^*,\mu^*)$ and eigenvector directions for $p<p_c$ when the fixed point is stable.

The eigenvector $u_1$  corresponds to a non-vertical direction along which many initial conditions converge to the fixed point, while the eigenvector $u_2$ simply the direction of axis $\rho$.

Taking into account Eq. (\ref{p_c}), Eq.  (\ref{l1l2}) can be written as 
\begin{equation}\label{l1l2new}
\lambda_1=\lambda_2\frac{1}{2}\left(1+\frac{1-p_c}{1-p}\right), 
\end{equation}
thus 
\begin{equation} \label{iff}
|\lambda_1|<|\lambda_2| \textit{ iff } p<p_c.
\end{equation}
In other words,  the eigenvector $u_1$ corresponds to the direction of slowest convergence to the fixed point in the space $(\rho, m)$.
%, which means that most trajectories approach fixed point from direction close to $u_1$.
As the parameter $p$ goes towards its critical value $p_c$, the convergence rates along both directions are similar and both tend to zero. 

\section{Numerical results}

Now we compare analytical results received from our mean-field approximation to numerical simulations of coevolving network of spins. When it is not otherwise stated we consider a network of $N=50000$ nodes, with mean degree $\mu=4$, where we observe variables ($\rho$, $n$, $m$). The critical value of the rewiring probability calculated from the mean-field approximation for such a system  is equal to $p^*=2/3$. To construct the network we set:

\begin{equation}
N_\alpha=N\frac{1+\alpha n}{2} 
\end{equation}

\begin{equation}
M_{\alpha\beta}=M_{\beta \alpha}=N\mu\frac{\rho}{2}
\end{equation}
\begin{equation}
M_{\alpha \alpha }=N\mu\frac{1-\rho+\alpha m}{2}
\end{equation}
Considering the above, we construct the network with $N_+$ nodes endowed with $+1$ spins randomly connected by  $\frac{M_{++}}{2}$ links and $N_-$ nodes endowed with $-1$ spins randomly connected by  $\frac{M_{--}}{2}$ links. Both groups are connected by $M_{+-}$ links. 

In a single update, a node $\textit{i}$ with a spin $s_i$ and one of its neighbors node $\textit{j}$ with a spin $s_j$ are chosen randomly. If $s_i=s_j$ nothing happens. Otherwise, with the probability $\textit{p}$, the link between $\textit{i}$ and $\textit{j}$ is reconnected from $\textit{j}$ to some randomly chosen node $\textit{l}$ such that $s_l=s_i$ or with the probability $(1-p)$ the spin $s_i$ is changing to $s_j$. 
Each time step consists of $N$ single node updates. Initial conditions are $\rho_0=0.3$, $n_0=0.2$ and $m_0=0.4$.

\begin{figure}[!hptb]
\begin{center}
\includegraphics[width=\columnwidth, angle=0]{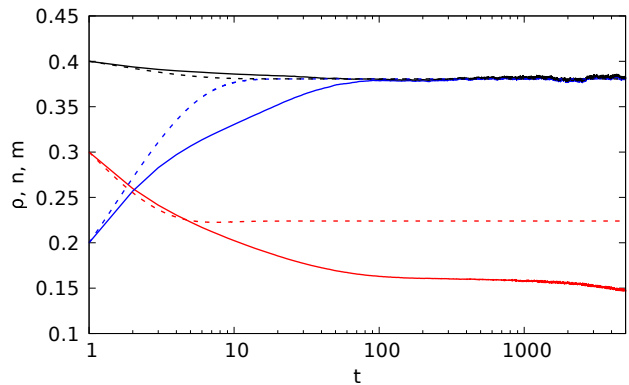}
\caption{(Color online)
Density of active links $\rho(t)$ (red lines)  does not drops to zero while the  mean magnetization of nodes $n(t)$ (blue lines) and links $m(t)$ (black lines) become equal when $p<p_c$. Data obtained from numerical simulations (solid lines) for the rewiring probability $p=0.3$ confirm analytical mean-field predictions (dashed lines). 
}
\label{r2}
\end{center}
\end{figure}

\begin{figure}[!hptb]
\begin{center}
\includegraphics[width=\columnwidth, angle=0]{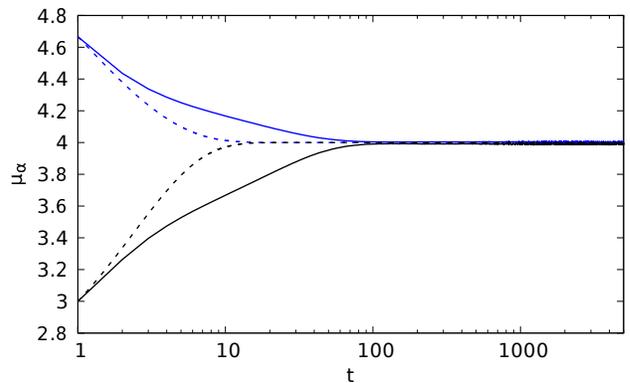}
\caption{(Color online)
Mean degrees of nodes  $\mu_\alpha$ endowed  with different spins $\alpha$ tend towards the same value during time evolution in active phase. Data obtained from numerical simulations (solid lines) for the rewiring probability $p=0.3$ confirm analytical mean-field predictions (dashed lines). The blue lines show $\mu_+(t)$ and the black lines show $\mu_-(t)$. 
}
\label{k2}
\end{center}
\end{figure}

\begin{figure}[!hptb]
\begin{center}
\includegraphics[width=\columnwidth, angle=0]{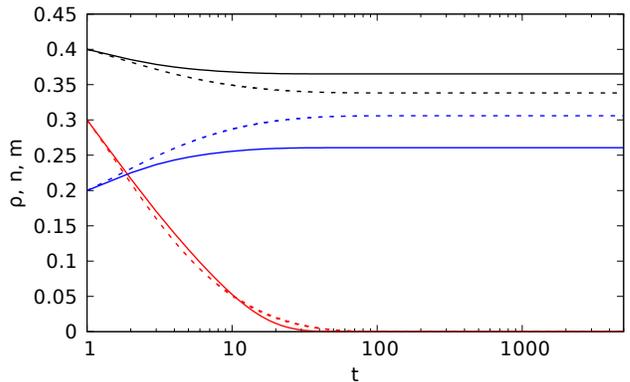}
\caption{(Color online) 
Density of active links $\rho(t)$ (red lines) drops to zero causing the mean magnetization of nodes $n(t)$ (blue lines) and links $m(t)$ (black lines) to freeze at different values if $p>p_c$. Data obtained from numerical simulations (solid lines) for the rewiring probability $p=0.7$ qualitatively confirm analytical mean-field predictions (dashed lines). 
}
\label{r1}
\end{center}
\end{figure}

\begin{figure}[!hptb]
\begin{center}
\includegraphics[width=\columnwidth, angle=0]{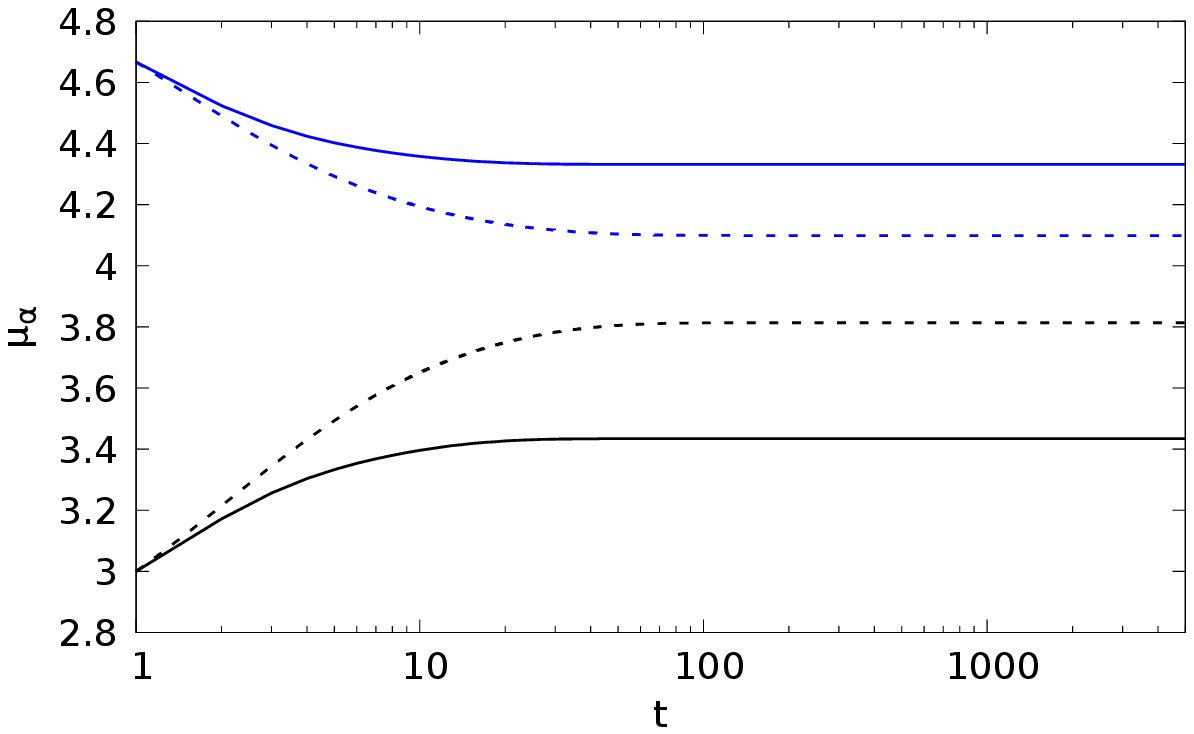}
\caption{(Color online) Mean degrees of nodes $\mu_\alpha$ endowed with different spins $\alpha$ do not become equal when $p>p_c$. Data obtained from numerical simulations (solid lines) for the rewiring probability $p=0.7$ qualitatively confirm analytical mean-field predictions (dashed lines). The blue lines show $\mu_+(t)$ and the black lines show $\mu_-(t)$. 
}
\label{k1}
\end{center}
\end{figure}

Analytical predictions based on our mean-field approach shown in Fig. \ref{r2} to \ref{k1} were obtained by solving differential equations \ref{derodete} to \ref{debetadete} with the classical Runge–Kutta fourth order method implemented in the R package deSolve \cite{deSolve}. The presented data from numerical simulations are averaged over 1000 realizations. 

\begin{figure}[!hptb]
\begin{center}
\includegraphics[width=\columnwidth, angle=0]{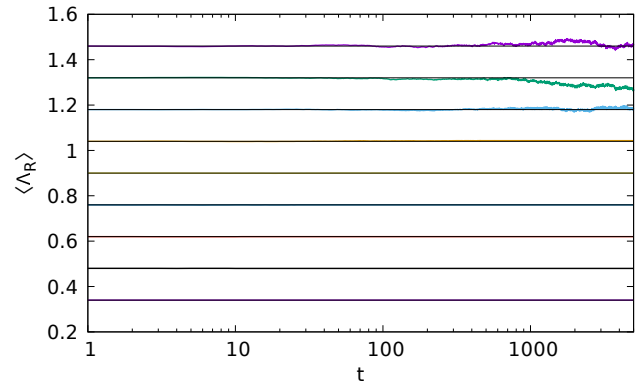}
\caption{(Color online) Numerical simulations of time evolution of the observable  $\langle \Lambda_R \rangle=(1-p) \mu \langle m(t)\rangle + p \langle n(t) \rangle$ [Eq. (\ref{cofmo})] confirm that in a very good approximation is a constant of motion.  Results are presented for  different values of the parameter $p$: 0.1, 0.2, ..., 0.9 (in order of decreasing $\langle \Lambda_R \rangle$). Horizontal black lines correspond to  theoretical values of $\Lambda$.
}
\label{x1}
\end{center}
\end{figure}

\begin{figure}[!hptb]
\begin{center}
\includegraphics[width=\columnwidth, angle=0]{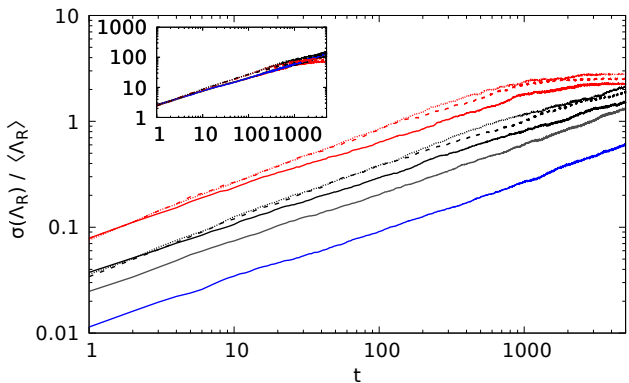}
\caption{(Color online) Standard deviation of the observable $\Lambda_R=(1-p) \mu m(t) + p n(t)$ (Eq. (\ref{cofmo})) divided by mean value of $\Lambda_R$ over different realizations increases in time as $\sqrt{t}$. 
Red lines present numerical data obtained from network with $N=1000$ nodes, black - $N=5000$, gray - $N=10000$, blue - $N=50000$. Solid lines correspond to  mean degree $\mu=4$, dashed lines to  $\mu=20$ and dotted lines to $\mu=40$.  \textit{Inset}: The same, except that ordinate values  are multiplied by $\sqrt{N}$ showing that standard deviation of $\Lambda_R$ decreases with size of the network as $1/\sqrt{N}$.
}
\label{std}
\end{center}
\end{figure}

The results at Fig. \ref{x1} confirm that for large networks, the value $\Lambda$ received  from our mean-field approach is indeed a statistical constant of motion, i.e., the conservation law given by Eq. (\ref{cofmo}) is fulfilled for averages over many realizations. 
For smaller $N$ we observed larger fluctuations, in particular, for small values of the probability $p$. These fluctuations can be explained through the finite size of the system. As can be seen in Fig. \ref{std} the standard deviation of the value of $\Lambda_R$ over realizations decreases with the size of the network approximately as $1/\sqrt{N}$. The standard deviation of $\Lambda_R$ reaches maximal value close to the consensus time, as individual realizations become fully ordered and so deviation approaches the maximum possible value (half realizations fully ordered to +, half fully ordered to -). It can be also observed that the growth of the tandard deviation in time follows $\delta(\Lambda_R) \sim \sqrt{t}$ which is characteristic to random-walk processes, showing that fluctuations of $\Lambda_R$ in particular realizations possess a  similar character to the random walk. \\
 Figsures \ref{r2} and \ref{k2} show simulations of the network when the probability $p$  of link rewiring is below $p_c$.  In such a case the density of active links connecting nodes of different spins does  not decay to zero in the course of time, what is seen at Fig. \ref{r2}. One can see in Fig. \ref{r2} that mean magnetizations of nodes and links tend toward equal values, i.e., $m=n$ in the stationary state and results of numerical simulations for their time evolution are in a qualitative agreement with analytical predictions. The difference between mean-field theory and the numerical calculations can be explained as an inaccuracy of our mean-field approximation as it disregards any correlations other than direct neighborhood, as well as possible statistical dependencies of node properties on degrees, such as was taken into account in \cite{demirel,pugliese}.
Results from Fig. \ref{k2} confirm also another prediction of mean field theory. In the course of time the presence of active links leads to the vanishing difference of degrees of nodes endowed with different spins, i.e.,  $\mu_+ = \mu_-$.        
Figures \ref{r1} and \ref{k1} show simulations of the network when the probability $p$ of the link rewiring is above $p_c$. In such a case the density of  active links connecting nodes of different spins  decays to zero in the course of time and  the difference between node magnetization  $n$ and link magnetization $m$ remains frozen what is seen  at Fig. \ref{r1}. Let us repeat that according to Eq. \ref{eq_mua} when $n \neq m$ then also $\mu_+ \neq \mu_-$ as can be observed  in Fig. \ref{k1}. 

\section{Discussion}

The main aim of this study is to understand the statistical relations between observables related to internal nodes variables (spins, opinions) and observables related to the network topology (link magnetizations) in the coevolving voter model.  Moreover we wanted to inspect if average   degrees of nodes are dependent on their spin values in such a system. To resolve these issues we extended the standard mean field approach by taking into account that during the system evolution mean degrees of nodes $\mu_{\alpha} = \langle k_{\alpha}\rangle$ with positive and negative spins as well as fractions of active links incident to them can be both different. Our analysis indicates that the link magnetization $m=(M_{++}-M_{--})/(N\mu)$ is coupled to the node magnetization $n=(N_+-N_-)/N$. Namely, a linear combination of these quantities forms a statistical constant of motion $\Lambda$ (Eq. (\ref{cofmo})), where  coefficients are the probabilities of spin flip and rewiring. In other words, there exists a special conservation law for node and link magnetization. Further, the difference $m-n$ decreases in time; if the asymptotic state is active, its mean magnetization of links equals to its mean magnetization of nodes  ($n^*=m^*$)   and in such a case the mean degrees of nodes do not depend on the spin value, i.e., $\mu_+=\mu_-$.  If there are  no active  links, i.e., the time evolution is frozen; the mean magnetization of links can be different from the mean magnetization of nodes, and the mean degrees of nodes is different for nodes of different spins.\\  

Numerical calculations confirm this picture in a short time scale, while in a longer scale fluctuations are visible.  

The fluctuations influence node and link  magnetizations in a different way. In the case where only the rewiring takes place ($p=1$), the node magnetization is exactly constant. On the other hand, if rewiring is absent ($p=0$), the link magnetization is constant only in the average. Hence the deviations of the solution from the mean field behavior are larger for small values of $p$, as observed in Fig. (\ref{x1}). The constant of motion $\Lambda$ complements the description in the full range of the probability $p$ between these two extrema.\\

The mean-field approach, where mean degrees of nodes in different states are treated as separate variables, is a generalization of previous calculations, where this distinction is not made \cite{ref_voter_coevolution, ref_voter_conservation}.
To place our result among other theoretical achievements on the voter model, we note that a large effort has been made recently to evaluate properly the fragmentation threshold $p_c$ where the active phase disappears \cite{demirel,ref_voter_fragmentationanalytical}. In particular, higher-order correlations between nodes of different opinions have been included to analytical modeling \cite{demirel,pugliese}. Our goal here is not to improve the accuracy of the mean-field approach of \cite{ref_voter_coevolution}, but to generalize it by including more degrees of freedom.

Applications of the coevolutionary voter model to the process of opinion formation were discussed in \cite{ref_wieland}, as often has been done with the voter model itself \cite{ref_castellano_review,ref_gracia}.
Our results should be useful for separating out the mean-field effects as well as for comparisons of the model results with real data on  social networks, where the number of neighbors of a person depends on her or his social status. Such a case can be observed  for co-existing communities representing, for example, various social classes \cite{Mann} where the density of social links can be different in various communities. Our study suggests that interactions between  such groups should reduce differences between these densities and a difference of  sizes of both groups should be proportional to a difference of number of their internal connections. We would like also to point out that in the case of  a strongly controversial  issue differentiating a society (e.g. abortion  or death penalty) the probability   of  acceptance  of another  opinion  can fall  below a critical value.  Then the social group separates into disjointed communities and  their structures quantified by the number of nearest neighbors  can be very different.       \\

The work was partially supported as  RENOIR Project by the European Union Horizon 2020 research and innovation program under the Marie Sk\l odowska-Curie Grant No. 691152 (project RENOIR) and by Ministry of Science and Higher Education (Poland), Grant No. 34/H2020/2016, No. 329025/PnH /2016. and National Science Centre, Poland Grant No.  2015/19/B/ST6/02612.
J.A.H. has been partially supported by the  Russian Scientific Foundation, Agreement No. \#17-71-30029 with co-financing of Bank Saint Petersburg and by a grant from The Netherlands Institute for Advanced Study in the Humanities and Social Sciences (NIAS).
\bibliography{bibliography}

\end{document}